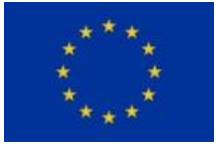
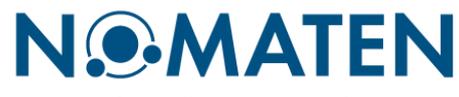

This work was carried out in whole or in part within the framework of the NOMATEN Centre of Excellence, supported from the European Union Horizon 2020 research and innovation program (Grant Agreement No. 857470) and from the European Regional Development Fund via the Foundation for Polish Science International Research Agenda PLUS program (Grant No. MAB PLUS/2018/8), and the Ministry of Science and Higher Education's initiative "Support for the Activities of Centers of Excellence Established in Poland under the Horizon 2020 Program" (agreement no. MEiN/2023/DIR/3795).

The version of record of this article, first published in Materials Science and Engineering: A, Volume 902, June 2024, 146590, is available online at Publisher's website: https://doi.org/10.1016/j.msea.2024.146590

# Nanoindentation responses of Fe–Cr alloys from room temperature to 600 °C

L. Kurpaska[1,*], M. Clozel[2], J. H. O'Connell[3], I. Jóźwik[1,4], E. Wyszkowska[1], W. Y. Huo[1,5,*], W. Chrominski[1,6], D. Kalita[1], S. T. Nori[1], F. Fang[7], J. Jagielski[1,4], J. H. Neethling[3]

[1] NOMATEN Centre of Excellence, National Centre for Nuclear Research, ul. A. Soltana 7, 05-400 Otwock, Poland
[2] German Aerospace Center (DLR), Institute of Material Physics in Space, Linder Höhe, 51147 Cologne, Germany
[3] Centre for High-Resolution Transmission Electron Microscopy, Nelson Mandela University, Port Elizabeth, 6031, South Africa
[4] Łukasiewicz Research Center - Institute of Electronic Materials Technology, Wólczyńska 133, 01-919 Warsaw, Poland
[5] College of Mechanical and Electrical Engineering, Nanjing Forestry University, 210037 Nanjing, China
[6] Faculty of Materials Science and Engineering, Warsaw University of Technology, Woloska 141, 02-507 Warsaw, Poland
[7] Jiangsu Key Laboratory of Advanced Metallic Materials, Southeast University, 211189 Nanjing, China

*Corresponding authors: lukasz.kurpaska@ncbj.gov.pl; wenyi.huo@ncbj.gov.pl

**Abstract**

In this work, the evolution of nanomechanical properties was studied systematically as a function of temperature, chemical, and microstructural complexity of different Fe-based alloys. Experiments were performed at different temperatures (room temperature, 200 °C, 400 °C, 600 °C) using the nanoindentation technique on low activation Fe9Cr–1WVTa (Eurofer97), model Fe-9Cr–NiSiP, Fe–9Cr alloys, and pure iron samples, followed by microstructural observations. The results show varying softening and hardening effects depending on the experimental temperature, demonstrating Portevin-Le-Chatelier effect, i.e., dynamic strain aging phenomenon in model alloys. Sources of the dynamic strain aging instabilities were traced back to the interaction between dislocations and alloying elements such as interstitial carbon and substitutional chromium. The materials undergo dynamic recovery and recrystallization below the regions of high-temperature indentation depending on the pre-indentation dislocation density and the alloy composition. Our findings help in the understanding of the structure and mechanical property relationship in complex Eurofer97 alloy at high-temperatures for potential nuclear applications as structural materials.

# 1. Introduction

Ferritic-martensitic steels (f-m) steels are a candidate material from which structural components of the next-generation reactors will be built. Among many different applications, they are planned to be used for the Molten Salt Reactor (MSR) and Lead Fast Reactor (LFR) [1-3]. However, according to the design specifications, these two reactor types will operate under very intensive neutron flux, and high temperature and will be cooled with highly corrosive media (e.g., liquid lead or sodium or lead/bismuth eutectics). Due to these coolants' fundamental thermodynamic and neutronic characteristics, these reactor concepts offer great potential for new reactor designs that achieve a high degree of inherent safety, simplified operation, and excellent economic performance while providing the fuel material management advantages characteristic of fast reactors. However, the operational conditions and reactor environment pose a challenge for prolonged usage of the materials from which structural components are built. One of the solutions to these problems is f-m steels. This is due to their reasonably good thermophysical and mechanical properties, reduced radiation-induced swelling and helium embrittlement under (fission) neutron irradiation, and largely good compatibility with cooling and breeding materials [2,4-7]. However, f-m steels are not perfect. They show problematic irradiation-induced embrittlement at low temperatures (< 350 °C), i.e., f-m steels lose ductility when irradiated at this temperature window. This phenomenon occurs even if the neutron dose is very low (according to Chen et al. [8], the effect saturates when a high neutron dose is used). Because of that, some constraints on the reactor design still exist, and they must take into account the period of the operation at lower temperatures or simply commonly known "cold spot" areas.

Since m-f steels have very complex microstructure, several authors pointed out that it is important to understand which material features are responsible for hardening and subsequent embrittlement in a wide temperature regime [5-11] (to explore their usage in different temperature operating windows). In addition to that, it was pointed out that the only way to do this successfully is to isolate the effect of temperature on particular features and progress with chemical complexity to account for the radiation damage in the final step.

The embrittlement effect of f-m steels in low temperatures has been analyzed as a function of the Cr content. Extensive studies commenced in the middle of the '90s, concluding that 9%Cr allows the ductile-brittle transition temperature (DBTT) shift to be minimized. Therefore, this Cr-content has been identified as optimal for f-m steels, which has been proven experimentally, while a physical explanation has been developed based on the modeling efforts [9]. Accelerated hardening of these steels (due to radiation defects) was described as to the Cr-content, which influences the microstructure evolution and features redistribution, which results in slowing down the interstitial loop motion [10], segregation around dislocation loops, and $\alpha$' precipitation [11-12].

Currently, a lot of European effort has been put into understanding the effect of radiation and temperature on the structural and mechanical properties of f-m steels [13]. Recent investigations on FeCrC model alloys containing minor additions of solutes such as Ni, Si, or P showed that solute clusters contain not only Cr but traces of these minor elements. These solute clusters form at low-temperature irradiation (300 °C), and the current understanding is that they are the main contributor to the radiation-induced hardening effect (demonstrated as yield strength increases) [14]. Usually, Cr-based NiSiP clusters are of size about 4 nm in diameter and are formed in similar density, about $10^{23}$ $m^{-3}$, regardless of the Cr-content (measurements have been performed for various Cr additions: 2.5, 5, 9, and 12%).

In conclusion, in m-f steels, not only the Cr content but also other factors were identified as potentially influencing the microstructure evolution under the simultaneous effect of temperature and irradiation, which leads to the subsequent hardening. This is also related to the effect of C-content in the matrix [15], which correlates with voids at lower Cr content and heterogeneous loop distribution for higher Cr content. Described complex microstructural changes, which evolve with the chemical composition of the material, must be understood in detail, taking into account the impact of temperature & irradiation independently, and both of these effects simultaneously, if one wants to explain mechanisms responsible for the hardening effect. For this reason, in the proposed research, we focus on the temperature effect and independently investigate the mechanical properties of the three model alloys and commercial Eurofer97 specimens, considering microstructural complexity progression. In addition to that, the quantity of the available material is very limited. Therefore, performing experimental tests using the nanoindentation technique and measuring the hardening effect has been chosen.

Nanoindentation is an important tool for measuring the hardness and elastic modulus of small volumes, irradiated layers, or coatings [16-20] and is also an emerging tool for studying fundamental physical phenomena. This is done by observing nanoscale mechanical instabilities and the movements of small populations of dislocations, which are responsible for the plastic deformation mechanism [21-23]. One such phenomenon is dynamic strain aging (DSA), which is macroscopically observed as the Portevin-Le Chatelier (PLC) effect, i.e., inhomogeneous deformation under stress. DSA is an interaction between the dislocations necessary to plastically deform the material and solute atoms [24-30]. It can occur according to two mechanisms: DSA due to interstitial atoms and DSA due to substitutional atoms. This phenomenon can also be separated according to the temperature zones in which it commences: (i) low temperatures where immobile solute atoms hinder dislocation movement (solid-solution strengthening), (ii) intermediate temperatures where the solutes are mobile enough to move to the dislocations and arrest their movement (jerky flow or serrations), and last, (iii) high temperatures where the solutes follow the dislocation movement, diffusing too quickly to hinder them. This results in a decrease in the mechanical strength of the material. The transition between these three temperature zones depends on the applied strain and/or dislocation velocity. This allows the activation energies related to the dislocation and solute atom interaction to be found. We observed the presence of these phenomena in pristine model Fe–9Cr alloys, and in this manuscript, we explain their impact on the reduction of the mechanical properties measured in situ at high temperatures by nanoindentation, simultaneously confronting this data with complex structural analysis.

The additional goal of this work is to prove that one can study the nanomechanical properties of ion-irradiated material in situ at high temperatures, thus investigating the hardening effect of these materials (triggered by radiation defects) at operational conditions of the power plant. Presented in this work, data were collected with a 50 mN load, resulting in deformation of about 0.75 to 1.25 um depth of the material (depending on the temperature and chemical composition of the studied material). Therefore, one can conclude that conducting in-situ high-temperature mechanical tests on ion-irradiated samples can be successfully done. Naturally, one needs to conduct irradiation with high-energy ions to develop a sufficiently thick irradiation zone to account for the size of the plastically deformed area below the indenter tip. In the following part of this study, we plan to investigate 10 MeV Fe-ion irradiated samples at the temperature of 300 °C.

This paper focuses on the high-temperature nanoindentation of four different model alloys with ferritic (pure Fe, Fe–9Cr, Fe9Cr–NiSiP) and martensitic-ferritic (Eurofer97) structures. Experiments were performed at room temperature (RT) followed by 200 °C, 400 °C, and 600 °C testing. After the tests, a comprehensive structural analysis was performed using scanning and transmission electron microscopes (SEM and TEM). Special attention has been put on grain size, orientation, and solute particle distribution. We analyzed mechanical data as a function of microstructural complexity progression. Recorded data prove the presence of the dynamic strain aging effect and its relation to the material's microstructure.

## 2. Experimental

### 2.1. Materials

The three model materials (Fe, Fe–9Cr, and Fe9Cr–NiSiP) used in this study were cast by OCAS (Gent, Belgium) in an induction vacuum furnace by additive melting. The pure Fe sample contained impurities of 50 ppm Co and 70 ppm Ni (all other elements were below 10 ppm). The Fe–9Cr sample contained 9.1 wt.% Cr and 0.0067 wt.% carbon. The chemical composition of the samples was measured using spark-source optical emission spectroscopy (SS-OES). This allowed the measurement of all elements except

Ni, Si, and Al. In addition, inductive coupled plasma-optical emission spectroscopy (ICP-OES) was used to measure the contents of Ni, Si, and Al. The chemical compositions are reported in Tab. 1. Elements that are not reported were below the detection limit. All model alloys were fabricated as a rolled plate with a final thickness of 10 mm. This was done by cutting from each lab cast a 50 mm × 125 mm × 250 mm piece. Then, these pieces were introduced to the preheated furnace at 1200 °C for 90 min. Following this step, hot rolling was performed (in 7 passes) until 10 mm was achieved. The hot rolling lasted for ~60 s, and the temperature of each element was approx. 930 °C after the termination of the process. Afterward, cooling to RT in the air was performed, resulting in a fully ferritic structure. The final dimensions of the sheets (measured in the rolling direction) were ~10 mm × 250 mm × 600 mm (height × width × length).

The Eurofer97 steel was produced by using the standard industrial procedure. The produced ingot was normalized at 1050 °C for 15 min, followed by water cooling. It is important to remember that the Fe, Fe−9Cr, and Fe9Cr−NiSiP samples have a fully ferritic matrix, while a ferritic-martensitic structure characterizes Eurofer97.

Table 1. Chemical composition of Fe, Fe−9Cr, Fe9Cr−NiSiP, and Eurofer97 specimens.

| Element (wt.%) | C | N* | Si* | P | S | Co | V | Cr | Mn |
|---|---|---|---|---|---|---|---|---|---|
| Fe | 0.006 | <0.005 | 0.001 | 0.003 | <0.005 | 0.005 | <0.001 | 0.002 | <0.001 |
| Fe-9Cr | 0.0067 | <0.005 | 0.004 | 0.003 | <0.005 | - | - | 9.1 | - |
| Fe-9Cr-NiSiP | 0.006 | <0.005 | 0.221 | 0.032 | <0.005 | - | - | 9.1 | - |
| Eurofer97 | 0.12 | 0.017 | 0.07 | <0.005 | 0.004 | - | 0.19 | 8.99 | 0.44 |

| Element (wt.%) | Ni | Ti | Mo | Cu | Ta | W | Al* | Fe |
|---|---|---|---|---|---|---|---|---|
| Fe | 0.007 | <0.001 | <0.001 | <0.001 | <0.001 | <0.001 | 0.023 | Bal. |
| Fe-9Cr | 0.009 | - | - | - | - | - | 0.027 | Bal. |
| Fe-9Cr-NiSiP | 0.092 | - | - | - | - | - | 0.028 | Bal. |
| Eurofer97 | 0.007 | 0.009 | <0.001 | 0.022 | 0.14 | 1.1 | 0.008 | Bal. |

*chemical composition checked with ICP-OES technique*
*The standard deviation was between 0.001 and 0.015 %, depending on the measured element. Elements with contents below 10 ppm could not be clearly identified by the measurement technique. Whenever questionable and for clarity, we left these columns empty.*

## 2.2. Sample preparation

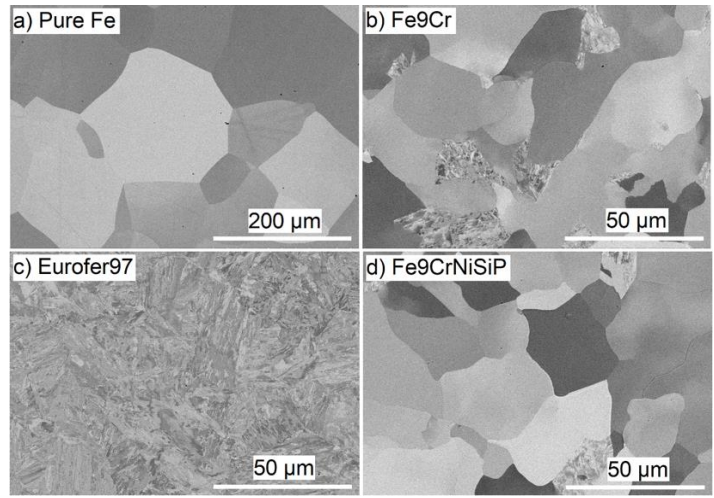

Figure 1. Microstructure of the materials: a) pure Fe, b) Fe–9Cr, c) Eurofer97, and d) Fe–9Cr–NiSiP observed under SEM.

The sample surfaces were all ground and polished on a MetaServ 250 (BUEHLER) in the following order: SiC grinding papers, from 600 down to 4000 grit, followed by the final polishing fluid – MasterPolish 0.05 µm (BUEHLER). The microstructures of all samples are shown in Fig. 1. It should be noted that the scale is different for pure Fe (Fig. 1a), as this material is characterized by large grains. Fe−9Cr and Fe−9Cr−NiSiP are similar in appearance, but Fe−9Cr presents more bainitic grains. All three materials, as shown in Fig. 1a, b, and d, are characterized by a single ferritic structure. Eurofer97, Fig.

1c) possesses a martensitic-ferritic microstructure and carbides dispersed on the grain boundaries and in the grains.

2.3. Nanoindentation

Nanoindentation tests were performed by using the NanoTest Vantage high-temperature device provided by Micro Materials Ltd. [31]. All experiments were conducted with a cBN Berkovitch indenter. Single-cycle nanoindentations were run in load-controlled mode: 10 s loading and 5 s unloading time. The dwell period was prolonged at elevated temperatures to 60 s to study the short-term creep behavior of the materials. The 60 s thermal drift measurement at the end of the indentation provides a measure of the displacement taking place during the measurement, which is solely due to the thermal disparity between the sample and the indenter. A correction is then applied based on this measurement when analyzing the results.

For each material, indentations were first performed at RT. Then, the temperature was increased separately at the indenter and in the sample holder until the temperatures at the sample surface and the indenter were matched. Afterward, the sample was brought in contact with the indenter, and the thermal disparity was calculated. Upon temperature stabilization, trial tests of 3 indents were run before each indentation to ensure that the thermal drift was acceptable. First, high-temperature indentations were made at 200 °C. Runs of 15 indentations were then performed after an additional thermalization time of 120 seconds while maintaining contact between the indenter and sample. The same process was repeated at 400 °C and then 600 °C. The procedure lasted ~72h for each sample, during which the chamber was constantly flushed with high-purity Ar to limit oxidation of the sample surface. The device is enclosed in a Plexiglas chamber [32], and a sensor measures the oxygen content in the chamber at all times. As presented in the following section, corrosion cannot be avoided and is one of the challenges when conducting high-temperature nanoindentation in controlled atmosphere conditions. However, as explained in the discussion section, the effect of corrosion is minimal due to sufficiently large indents. No indentations were performed at 600 °C on pure Fe, as the sample suddenly presented a large thermal disparity between the sample surface and the indenter. At this temperature, oxidation was considerable enough to prevent further indentations (further explanations in the following section). Following the indentations at the highest temperature (600 °C), the system was allowed to cool down, and the sample was removed from the sample holder to be taken for microstructural observations.

In addition to limiting the surface roughness when using nanoindentation, it is also essential to limit the thermal disparity between the sample surface and the indenter as much as possible. Our setup allows us to control 3 temperatures independently: the indenter, the stove on which the sample is mounted, and the sample surface where the measurement is taken. To minimize thermal drift during the experiment, the temperature adjustment is always done between the indenter and the sample surface, which is done by adjusting the stove temperature.

### 2.4. Microstructural observations

SEM was conducted on a Carl Zeiss Auriga on all samples after indentation. To observe more closely how the microstructure of each material responded to nanoindentation deformation, TEM was conducted on a JEOL ARM 200F at the Centre for High-resolution TEM of Nelson Mandela University. FIB lamellae were taken from indent cross-sections using an FEI Helios NanoLab 650 system. This was done after indentations at RT and 600°C. Transmission Kikuchi diffraction (TKD) was performed on the TEM-ready lamellae in a JEOL JSM7001F SEM equipped with an HKL Nordlys camera.

## 3. Results

### 3.1. Nanoindentation

The results of high-temperature nanoindentation are presented in Figs. 2 to 5. First, the typical load-displacement curves recorded at different temperatures for each material are shown in Fig. 2. Displacement bursts marked with black arrows are observed for pure Fe (Fig. 2a) at 200 °C, Fe–9Cr (Fig. 2b) at 400 °C and Fe–9Cr–NiSiP (Fig. 2c) at 400 °C. No bursts were observed for Eurofer97. To further examine the presence or absence of these bursts, the derivatives of the displacement-time curves are presented in Fig. 3. One can observe that the bursts are present in the form of plain peaks. In the case of pure Fe, they appear at 200 °C only, and the bursts on the second half of the curves seem to coincide. This effect is not clearly visible for Fe–9Cr and Fe–9Cr–NiSiP and appears more randomly. No bursts are visible in the case of Eurofer97, which is demonstrated by the recondense of smooth load-displacement curves (Fig. 2d) and lack of sharp peaks when plotting derivatives (see Fig. 3, bottom line).

We conducted indentations with different loads to understand whether the appearance of the bursts is a load-dependent process. The experiment was performed on Fe–9Cr specimens at 25 mN, 50 mN, 75 mN, and 100 mN loads. See Fig. A in the supplementary material where we compared the derivatives obtained from this experiment. One can observe that the higher the load is, the higher and more numerous the peaks, which proves that the recorded phenomenon is load-dependent and that 50 mN force is enough to record described bursts.

The hardness results as a function of temperature are shown in Fig. 4a, while the Young modulus plot is presented in Fig. 4b. The error bars represent standard deviations calculated from circa. 7-10 data points. It was expected that higher temperatures would entail lower hardness. However, at 400 °C for the Fe-9Cr and Fe-9Cr-NiSiP specimens (where we recorded burst events), we observed an increase in hardness and Young modulus. At the same time, the Eurofer97 specimen shows a decrease in both parameters. The pure Fe sample presents a similar effect to Fe-9Cr and Fe-9Cr-NiSiP, yet at 200 °C. Again, the hardening effect is accompanied by displacement bursts (see Fig. 3). After recording the first displacement bursts for a given temperature, the subsequent increase in temperature greatly reduces the hardness in all materials, and the burst disappears.

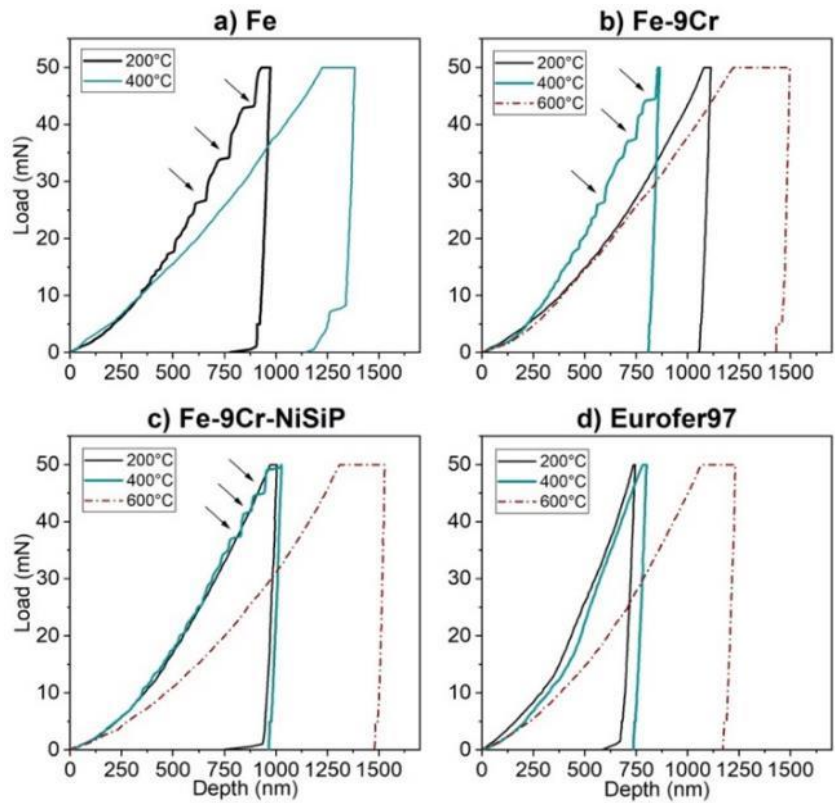

Figure 2. Typical load-displacement curves at elevated temperatures for all materials.

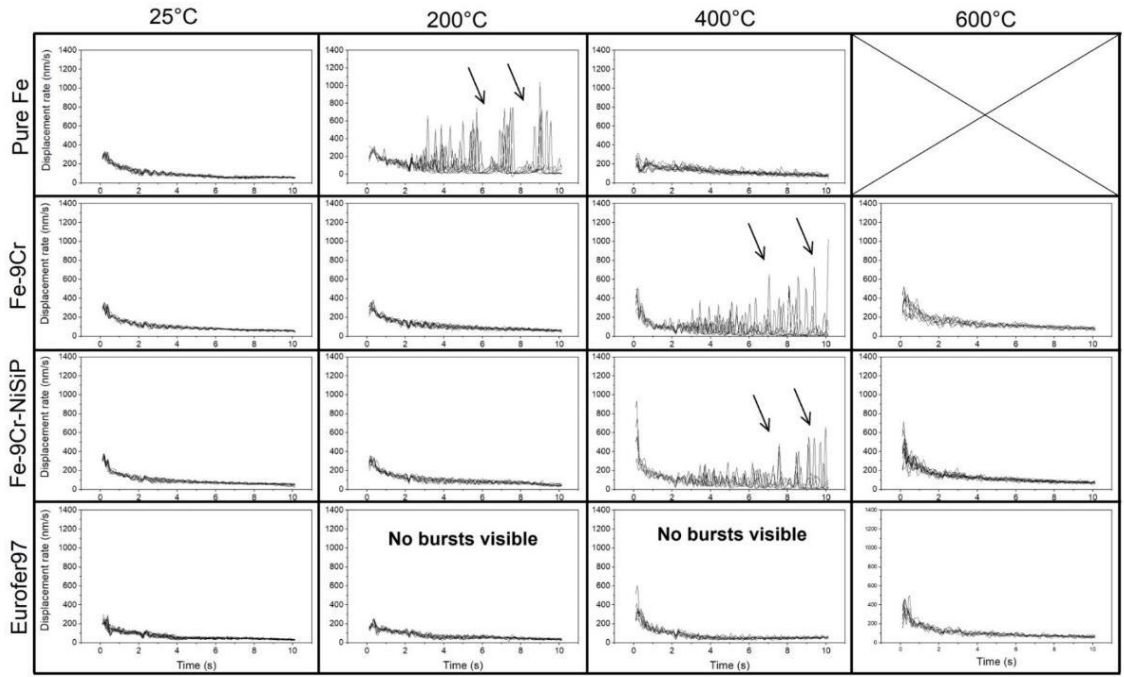

Figure 3. Derivatives of the displacement-time curves for different temperatures recorded for 50 mN indentations. Arrows point to the presence of the burst.

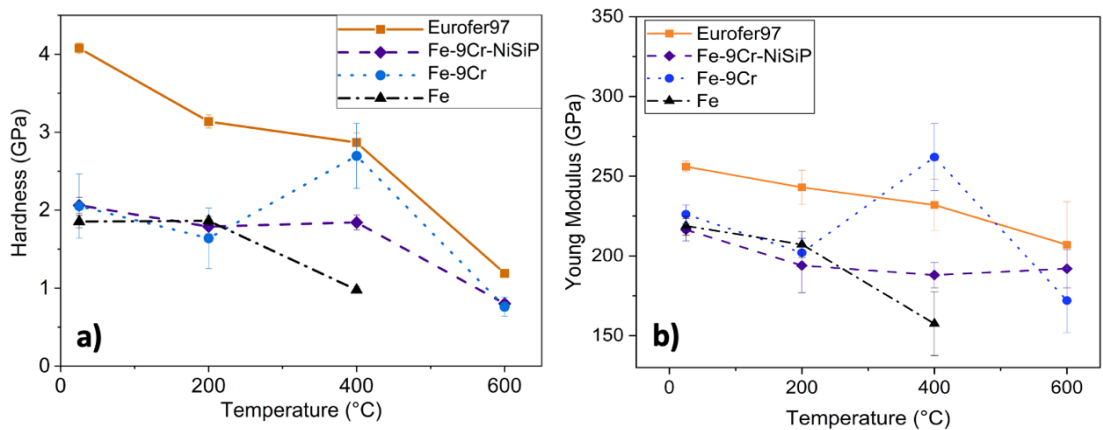

Figure 4. Effect of temperature on (a) indentation hardness and (b) Young modulus at 50 mN loads for all studied materials.

To better understand the behavior of these materials at high temperatures, the short-term creep was measured by holding the indenter at a maximum load (50 mN) for 60 s. Fig. 5 shows the depth vs. time curves recorded during the holding period. The Fe−9Cr sample

(Fig. 5b) presents the highest standard deviation. More substantial variations in results have also been noticed with this material at RT. We believe that the source of this instability is related to the presence of bainitic grains within the ferritic matrix, as shown in Fig. 1b). The slopes of the short-term creep are shown in the graphs, where all creep curves were superposed into an average curve and deviation. At 200 °C and 400 °C, all samples are considered to have reached a constant creep slope during the 60 s holding time (before unloading started). Whether this was the case at 600 °C is less certain, especially in the case of Fe–9Cr and probably Fe–9Cr–NiSiP. These samples showed the highest – and still increasing – standard deviation of creep.

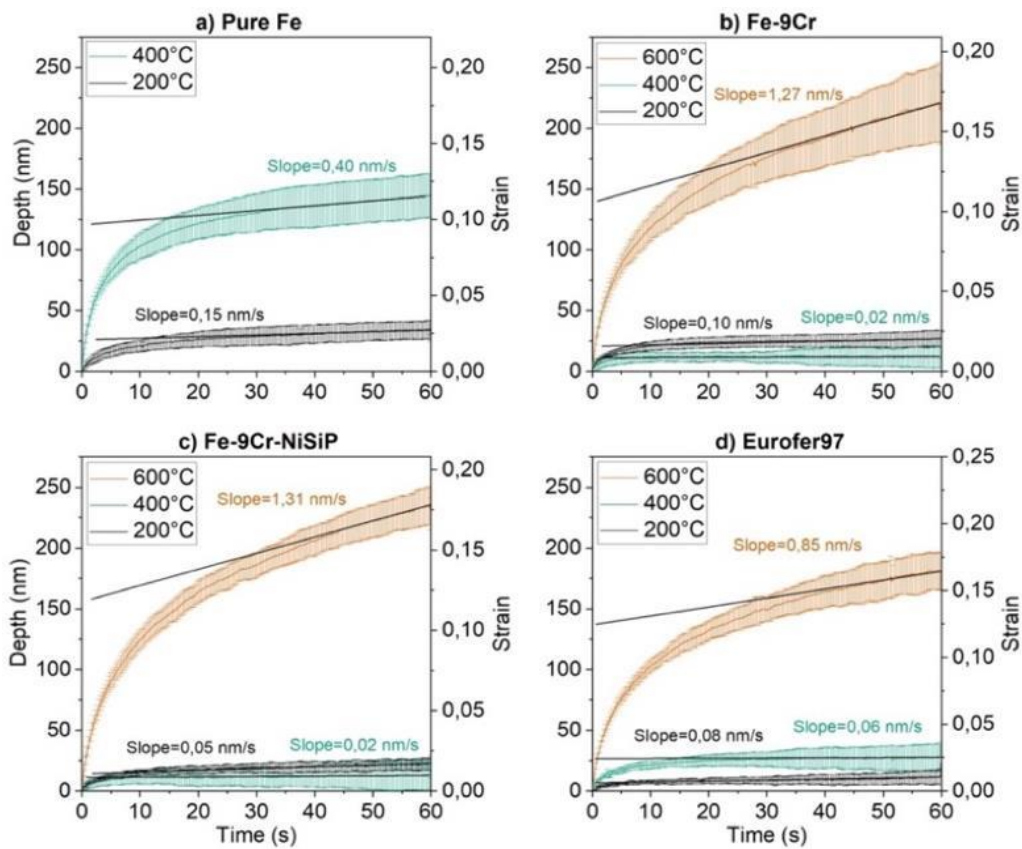

Figure 5. Creep measurements recorded over 60 s under a maximum load of 50 mN for all materials.

### 3.2. STEM observation

The low-angle annular dark field (LAADF) STEM images of the indent cross-sections are presented in Fig. 6. The specimen surface is shown at the top of the image in all cases. The asymmetry in the indent profiles is purely due to the cross-sectional cutting with FIB of a 3-fold symmetric indent. The steep slope is the flat face, and the other is the valley along the opposite edge of the indenter tip.

In the case of RT indentation of the samples pure Fe and Fe-9Cr (Fig. 6 a,b), the dislocation networks are concentrated on either side of the indentation tip. Such localization of the dislocations eases out in the case of Fe–9Cr-NiSiP and Eurofer97 specimens (Fig. 6 c,d), where dislocation forest networks are observed. In addition, Figs. 6 a-d also show that dislocations form cell structures near the indenter tip with a higher tendency in pure Fe. The tendency of dislocation cell formation reduces as chemical complexity increases for Fe-9Cr, Fe-9Cr-NiSiP, and Eurofer97. At high temperature (HT) (600°C), the dislocations arrange to form polygon structures for Fe-9Cr and Fer-9Cr-NiSiP alloys below the indent. In the case of Eurofer97, similar dislocation arrangements are observed, however, within the grains. The depth into which the indentation-induced dislocations extend within the sample would depend on the grain orientation and presence of grain boundaries (as visible in Fig. 6g). This will be examined in more detail in future work, while the detailed explanation of the occurring phenomena is given in the discussion part.

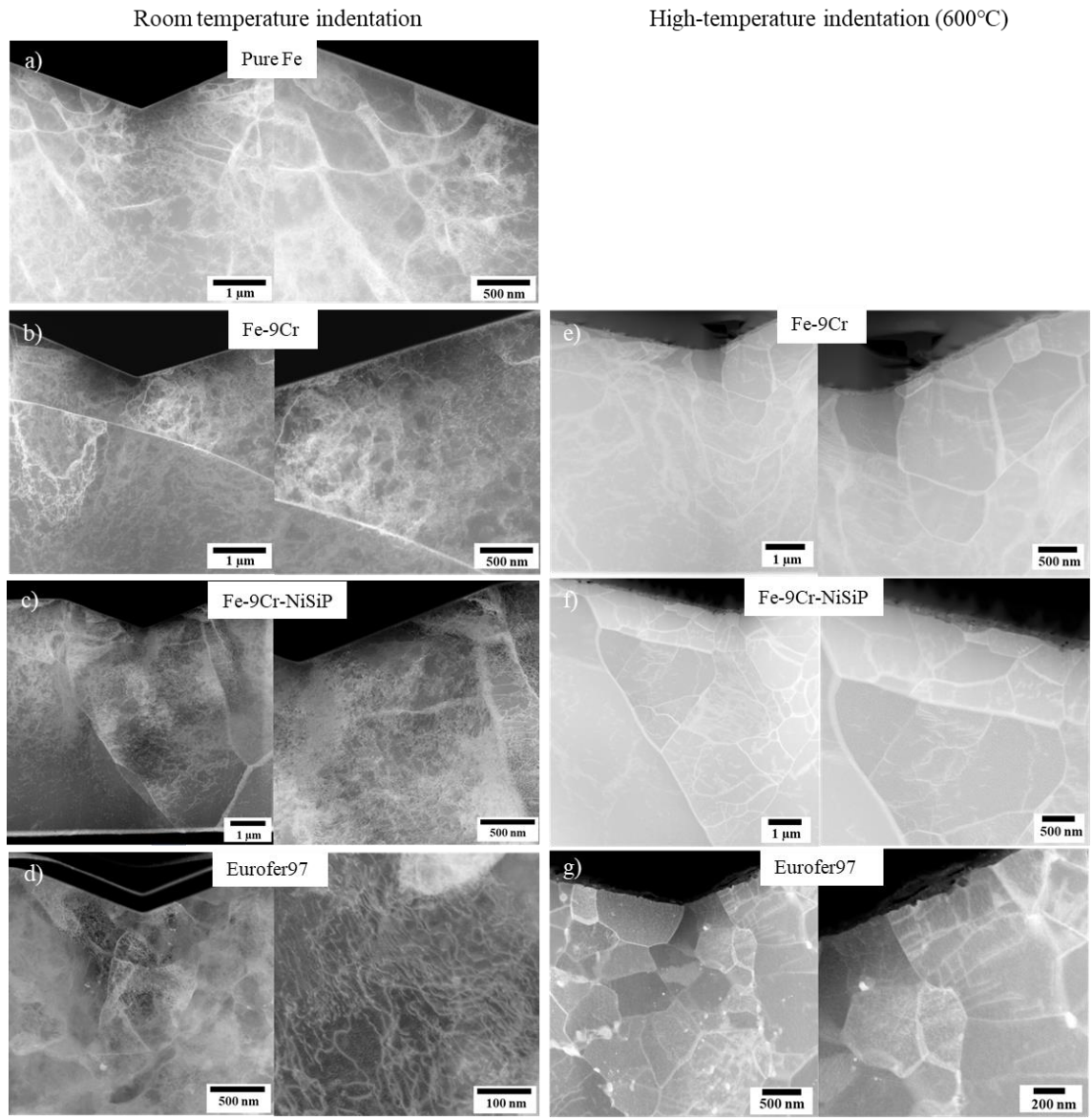

Figure 6. LAADF observations of all materials' cross-sectional indents (cut using the FIB lift-out technique). Comparison between indents made at room temperature and 600 °C. The absence of the pure Fe indent at 600 °C is related to the rapid oxidation of the system at this temperature.

Fig. 7 shows band contrast maps and inverse pole-figure (IPF) color-coded orientation maps of indentations made at RT and 600 °C in Fe–9Cr, Fe–9Cr–NiSiP, and Eurofer97 steel samples captured via TKD. The band contrast maps represent the quality of the Kikuchi bands detected by the EBSD camera. Regions around grain boundaries show overlapping patterns with typically poor signal-to-noise ratios, causing them to appear darker in the image. The stress field around dislocations also degrades pattern quality, but to a lesser extent, and thus, they are not as clearly seen as in the LAADF images. This makes the band contrast maps very useful for studying the grain structure.

From the band contrast and orientation maps in Fig. 7 and LAADF images in Fig. 6, it is clear that RT indents produce denser dislocation networks near the indent; hence, they show higher strain fields near the indent, specifically for Fe, Fe-9Cr, and Fe-9Cr-NiSiP. For Eurofer97, the RT indentation band contrast and IPF maps illustrate elongated grains due to the pristine microstructure along with higher strain fields near the indent. For instance, some small grains around the indentation faces with some intra-grain strain are visible in the IPF map, especially near the indent, but also to a lesser degree within grains that are relatively far away. For 600 °C indentation, as shown in Fig. 7 and confirmed by Fig. 6, the dislocations substructures arrange themselves to undergo polygonization for Fe, Fe-9Cr, and Fe-9Cr-NiSiP samples. In the case of Eurofer97, Fig. 7 shows grains exhibit relatively lesser strain levels compared to the RT indentation. Also, the elongated grains observed in RT indentation transform into equiaxed grains post HT indentation. Table 2 shows the grain size information of all the samples calculated after RT and HT indentations, which clearly indicates grain recrystallization after HT indentation for Eurofer97, demonstrating the transformation of elongated grains to equiaxed grains.

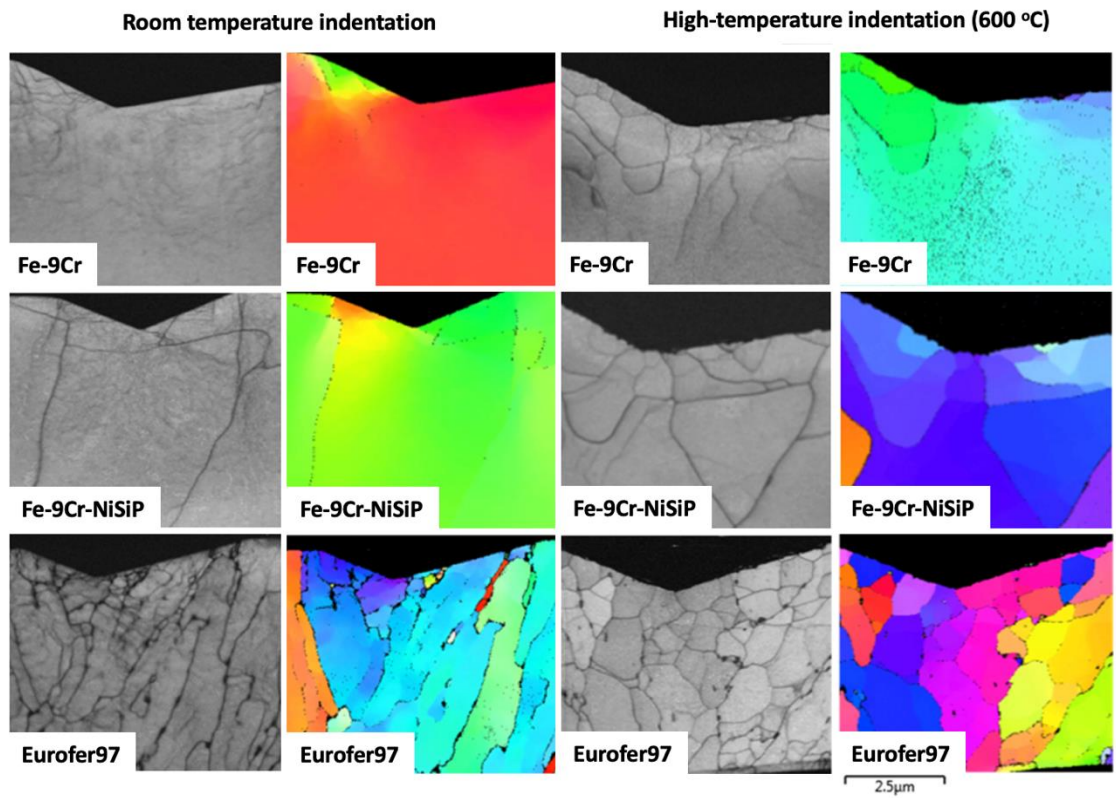

Figure 7. TKD band contrast maps (black-grey images) and IPF orientation maps (colored images) of indents made at RT and 600 °C in Fe–9Cr, Fe–9Cr–NiSiP, and Eurofer97 steel.

Table 2. Grain sizes of the samples in the regions below the indent at room temperature and high temperature (600°C)

| Grain size | Pure Fe | Fe-9Cr | Fe-9Cr-NiSiP | Eurofer97 |
|---|---|---|---|---|
| Below indent at the room temperature | mono | mono | 9 μm | 6.73 μm |
| Below indent at the high temperature |  | 7.4 μm | 11.2 μm | 5.2 μm |

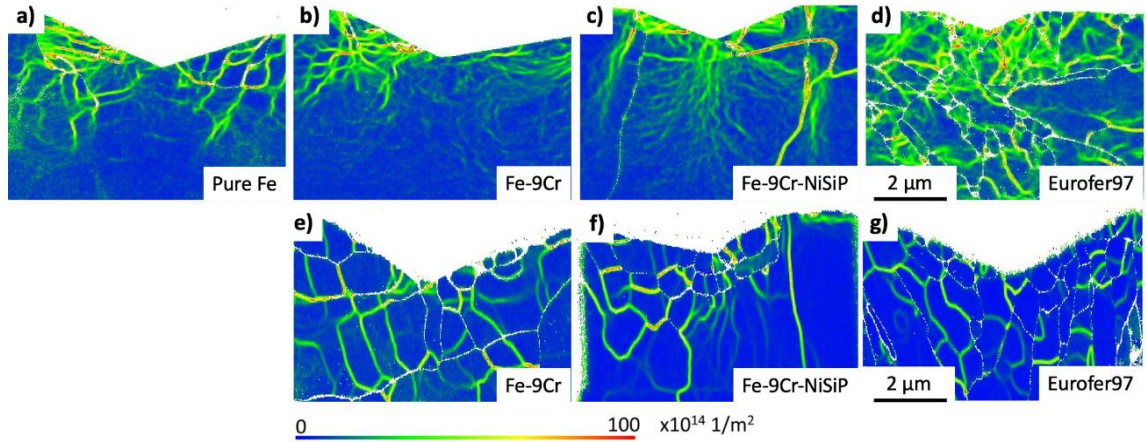

Figure 8. Density of the geometrically necessary dislocations calculated below the indents made at room temperature and after indentations at high temperature. Calculations were made based on TKD imaging using the approach proposed by Konijnenberg [33].

Table 3. Dislocation densities of the samples in the pristine region and the regions below the indent at room temperature and high temperature (600°C)

| Dislocation density [1/m$^2$] | Pure Fe | Fe-9Cr | Fe-9Cr-NiSiP | Eurofer97 |
|---|---|---|---|---|
| Pristine sample (no deformation) | 0.61 x 10$^{12}$ | 8.61 x 10$^{12}$ | 12.09 x 10$^{12}$ | 95.59 x 10$^{12}$ |
| Below indent at the room temperature | 12.67 x 10$^{14}$ | 11.39 x 10$^{14}$ | 12.67 x 10$^{14}$ | 24.65 x 10$^{14}$ |
| Below indent at the high temperature | | 13.24 x 10$^{14}$ | 12.7 x 10$^{14}$ | 15.27 x 10$^{14}$ |

To further understand the information presented in Fig. 6 and 7, we show dislocation densities below the indent calculated based on TKD imaging in Fig. 8. We used an approach proposed by Konijnenberg [33]. For these calculations, we applied Channel5 software. Kernel average misorientation on a square grid has been employed for estimating orientation variances between pixels. Then, the number of dislocations responsible for these variances was calculated and plotted. Blue indicates a low density of dislocations, while yellow and red depict the highest number of dislocations. The spatial dislocation distributions for the Fe, Fe–9Cr, and Fe–9Cr–NiSiP specimens at RT are similar, as illustrated in Fig. 8 and Table 3.

Table 3 describes the dislocation densities of the samples from the pristine regions and the regions below the indents for both RT and high-temperature cases. For the pristine condition, pure Fe contains the lowest density of dislocations, with one order more for Fe-9Cr and Fe-9Cr-NiSiP and one order more for Eurofer97. After RT indentation, the samples show two orders of improvement in dislocation densities. While high-temperature indentation led to minimal change in dislocation densities for Fe-9Cr and Fe-9Cr-NiSiP, Eurofer97 showed a significant reduction (~38%) compared to RT indentation.

As shown in Fig. 8, one can observe the tendency that as the alloying element content increases, the area directly beneath the indent tip seems to present more dislocations, see Fig. 8a-c. This is not the case for the Eurofer97 specimen, which shows dislocations distributed in the grains and grain boundaries for RT measurement, see Fig. 8d. This is most likely related to the manufacturing method. After indentation at 600 °C, Fig. 8e-g, the dislocation density is lower within the grains than at RT.

A closer look at the TEM results of pure Fe (see supplementary material Fig. B) points to the conclusion that the bulk of the deformation takes place on the sides of the indent. Dense dislocation networks confirm this in those areas (marked with red circles). The area between those circles is mostly bare but does present some dislocations. One can see that the dislocation density decreases with increasing distance from the surface. This result seems to be in agreement with the work of McLaughlin and Clegg [34], where misorientation was measured beneath indents in copper. The largest misorientation angles were reported to appear on both sides of the indent. This is in the same location as the areas rich in dislocations in the case of the pure Fe specimen.

## 4. Discussion

The influence of dynamic strain aging is very well seen in both Fig. 5a and b, where for the pure iron specimen (at 200 °C) and Fe-9Cr (at 400 °C), one can observe Hardness (Fig. 4a) and Young modulus (Fig. 4b) increase. One can also observe that the creep rate suspends at these temperatures, which is favorable for the dynamic strain aging to appear, see Fig. 5a - b) – green color-coded slope. Later, the opposite trend is recorded, as hardness is significantly affected by the DSA in the approximate 200 °C to 400 °C, for pure iron and Fe-9Cr, respectively. The normal temperature-depended hardness trend is recovered at RT and 400 °C, almost linearly reducing from ~1.8 GPa to ~1.0 GPa. On the other hand, the DSA-affected values reach almost 1.9 GPa at 200°C for pure iron and ~2.8 GPa at 400 °C for Fe-9Cr specimen, which is not anticipated if carbon content is lower. Nevertheless, the obtained Young's modulus values are presented in Fig. 4b and compared to the ones of Eurofer97 (obtained using E111-04 ASTM standard), matches well and follows the expected trend of decreasing as temperature grows, although Eurofer97 is stiffer at lower temperatures. This behavior might be expected due to characteristic microstructural features present in steel and alloying with 9% of Cr and other minor elements.

One can observe that the material's elastic modulus does not agree well with the commonly established value (~210 GPa) measured by other means, e.g., tensile test. We expect that the unloading stiffness $S$ has been most likely affected by indentation pile-up formation, which is a well-known problem [37]. As the testing temperature grows, the measured values tend to approach the expected value of Young's modulus. This is another confirmation, as the indentation pile-up formation is known and found to decrease with the ambient temperature. We expect that stiffness can also affect recorded hardness calculations to some extent. In the forthcoming contribution, the stiffness correction procedure will be applied [38]; however, first, true young modulus values calculated based on the tensile test would have to be found for each material (according to the author's knowledge, this data has been calculated to pure iron and commercial Eurofer97, but not for investigated two model alloys).

Fig. 3 presents the recorded displacement rate over time during the 10 s of the loading process. One can observe that the background noise is particularly elevated in the first 2 s of loading. It diminishes significantly after that, finally stabilizing in the second part of the loading process. This is likely to appear due to the time resolution of the device in the depth-time curves. However, recorded deviations (arising as a sudden increase of the displacement rate over time) observed in pure Fe, Fe–9Cr, and Fe–9Cr–NiSiP (marked with arrows) are clear. In the literature, this effect is called displacement burst (with several occurring effects that play a role simultaneously). The appearance of these bursts is also correlated with the presence of the hardening effect (see Fig. 4). The bursts may have their origin in Snoek relaxation and/or dynamic strain aging (DSA), dislocation nucleation and/or multiplication, the formation of precipitates, or finally breaking of an oxide layer (however this process is unlikely, as the oxide layer according to our studies is much less than 100 nm thick and bursts appear at a depth of 500 nm and more, pointing to the conclusion that recorded signal originates from a few μm which were deformed). Similar hardening at high temperatures has been observed in [39] and was correlated with the effect of oxidation. However, our nanoindentation has been performed in a protective atmosphere [23], which limits the oxidation rate. In conclusion, in our opinion, recorded hardening cannot be due to oxidation. Apart from Eurofer97, where many precipitates are present in the as-received material, no precipitates are visible in Fe–9Cr and Fe–9Cr–NiSiP at 600 °C. If they were the cause of the displacement bursts at 400 °C, they would have had to dissolve by the time the samples reached 600 °C, which again seems unlikely. According to the literature [29-30, 40-41], these bursts and the sustained hardness levels stem from the DSA effect. Using this explanation, one may distinguish three temperature ranges for each material:

- Low temperatures where dislocation movement is impaired by immobile solute atoms (solute-hardening);
- Medium temperatures where the solute atoms are sufficiently mobile to diffuse to nearby dislocations, causing pinning and thereby limiting deformation (hardening): this is where DSA occurs;
- High temperatures where the diffusion is increased further, allowing the solute atoms to follow dislocations (softening).

In our studies, pure Fe displayed DSA at ~200 °C, while Fe–9Cr and Fe–9Cr–NiSiP displayed DSA at ~400 °C. This can be explained in the following manner:

- In pure Fe, DSA at ~200 °C is recorded due to the presence of carbon (and nitrogen) atoms which diffuse to the dislocations. According to the chemical composition of the specimens, see Tab. 1, the pure Fe specimen contains 0.006 wt.% of carbon and less than 0.005 wt.% of nitrogen. Caillard et al. [29, 30, 40, 41] studied this effect extensively and proved that the presence of even 1 ppm [29] of C (in the second case, 16 ppm of C was studied [30]) is enough to block dislocation progress and impair their mobility. Caillard suggested that the dynamic interaction between carbon and mobile dislocations should take place up to 300 °C and is manifested by jerky flow (for lower temperatures) and serrated flow (for higher temperatures) and corresponds to the conventional Cottrell interaction between carbon and dislocations. Our work confirms this finding as we detected DSA at 200 °C via recorded bursts, but no bursts were detected at 400 °C. Moreover, the presence of nitrogen with an amount as low as 45 ppm was reported by Pink et al. [42] as being sufficient to induce strain aging in steels, further supporting our statement that the strain aging phenomenon appears in pure iron.

- The bursts do not appear in Fe−9Cr and Fe−9Cr−NiSiP at 200 °C – even though these samples contain traces of carbon and a significant amount of chromium content. Chromium is known to have a stronger affinity to carbon than Fe [29], so Cr atoms "capture" the C atoms, and these CrC complexes diffuse to the dislocations when the temperature is sufficiently high – approximately 400 °C. This is the main reason why this effect is delayed in the case of these two materials. Similar bursts were observed in a Fe−5Cr with 16 ppm carbon during *in-situ* straining in TEM at 400 °C [23, 29, 39]. Once again, increasing the temperature above 500 °C resulted in the disappearance of the bursts for the Fe−5Cr model material. Furthermore, it has been reported that no interstitial carbon was detected for Fe−Cr−C−X alloys with Cr contents above 2.5% [43]. The reported result once again confirms the literature data, i.e., the absence of bursts in the Fe−Cr alloys at 200 °C is related to C trapping by Cr.

The absence of DSA in the case of Eurofer97 is probably due to the martensitic microstructure and the presence of solute atoms, which increase the solution hardening. In addition, Eurofer97 is characterized by a dislocation-rich microstructure with the addition of carbides. One must remember that additional solute atoms provide barriers for mobile dislocations. Therefore, one would expect hardening as solutes should block dislocation movement. However, the data that was obtained show that nanoindentation, even at 600 °C, is still too low in the case of Eurofer97 to initiate displacement bursts. This suggests the presence of another mechanism responsible for the plastic deformation of this material.

There are several possible explanations for the difference in hardening observed between Fe−9Cr and Fe−9Cr−NiSiP. It is unlikely that Fe−9Cr and Fe−9Cr−NiSiP would display a peak in DSA at the same temperature, so Fe-9Cr may increase more in hardness than Fe−9Cr−NiSiP in the results because 400 °C is closer to the temperature at which DSA of Fe−9Cr reaches its peak than to that of Fe−9Cr−NiSiP. It is also likely that because of the additional impurities/alloying elements, Fe−9Cr−NiSiP is closer in behavior to Eurofer97, which experienced only a slight decrease (as opposed to an increase for the other materials) in hardness at 400 °C. Increased piling-up during nanoindentation of Fe−9Cr is also a possibility. Previous results reported by Malerba et al. [44] and Terentyev et al. [45] suggest that increased hardening in ion or neutron-irradiated materials is due to the presence of Cr, which tends to accumulate around dislocations affecting their glide. In contrast, the same team recently postulated that the presence of minor solutes such as Ni, Si, and P is essential for microstructural features (most likely solute-rich clusters) to be an efficient obstacle to dislocation motion [46]. In conclusion, the hardening mechanism of m-f steels remains unclear, and further investigation should be performed to test different hypotheses.

The periodicity and synchronicity of the later bursts in pure Fe are intriguing. The synchronicity may suggest that there is a spatial factor involved. In contrast, the periodicity suggests the possibility of a constant energy barrier needing to be overcome for dislocations to move and, therefore, for plastic deformation to take place because mostly C is present in pure Fe. Meanwhile, Cr, Ni, Si, and P are present (in varying quantities) in Fe–9Cr and Fe–9Cr–NiSiP, possessing different sizes and diffusion properties, possibly leading to barriers with many different energies.

Bearing that the formation of new precipitates or solute segregation during high-temperature nanoindentation may significantly affect the measured mechanical properties of the studied materials, detailed TEM studies on these phenomena were conducted. Fig. 9 shows the STEM micrographs and corresponding energy dispersive spectroscopy (EDS) elemental maps recorded for the Eurofer97 samples after nanoindentation at RT and 600°C. In both samples, relatively large Cr- and W-enriched particles were found mainly at the grain boundaries. Considering their chemical composition and the previous study on the precipitation behavior in Eurofer97 (e.g., [47]), they are identified as $M_{23}C_6$ carbides. The performed analysis also revealed the occurrence of two distinct types of finer particles enriched with V and Ta. The first type exhibited a nitrogen enrichment, indicating they are vanadium nitrides (VN). The second type corresponds to tantalum carbides (TaC). Both of them were previously reported in Eurofer97 [48]. The results show that annealing of Eurofer97 at a temperature of 600°C during the nanoindentation did not affect the occurrence of the strengthening particles. Therefore, the precipitation process should not affect the measured nanomechanical properties in this case. It is important to note that further study on elemental distribution within the matrix (with a special focus on the grain boundaries – see supplementary materials describing TEM/EDS line scan in Fig. C) does not reveal any sign of solute segregation in Eurofer97 at the temperature of 600 °C.

However, a slightly different behavior was observed for the Fe-9Cr-NiSiP case (see supplementary materials Fig. D). One can see traces of Cr segregation at the grain boundary. The observed phenomenon could be related to the vacancy diffusion mechanism [49-52]. However, one knows that this depends on several factors, such as grain boundary misorientation [53], temperature [49], or (if the material works in the nuclear environment) irradiation conditions [50, 51]. It has been proved that high-angle grain boundaries (HAGBs) are more prone to segregation [52] than low-angle grain boundaries (LAGBs). This different behavior is attributed to the variations in dislocation density at different grain boundaries. As reported by Xia et al. [53], Cr segregation at the grain boundaries may occur over a relatively wide temperature range (300 - 640°C). However, the process disappears at temperatures higher than 640°C, due to a change of the Cr diffusion mechanism. The misorientation angle in our case was about 5°, which indicated that the tests were done on the low-angle grain boundary. Hence, in this case, only the fine enrichment of Cr in the vicinity of the grain boundary is observed, as shown using the EDS technique (see supplementary material Fig. D).

Regarding the lack of segregation in the case of Eurofer97, this effect is related to the duration of the experiment. Cr segregation in these alloys has been reported only after prolog thermal treatments, exceeding 1000h [54] or due to the combined effect of very high damage (over 70 dpa) and temperature [55].

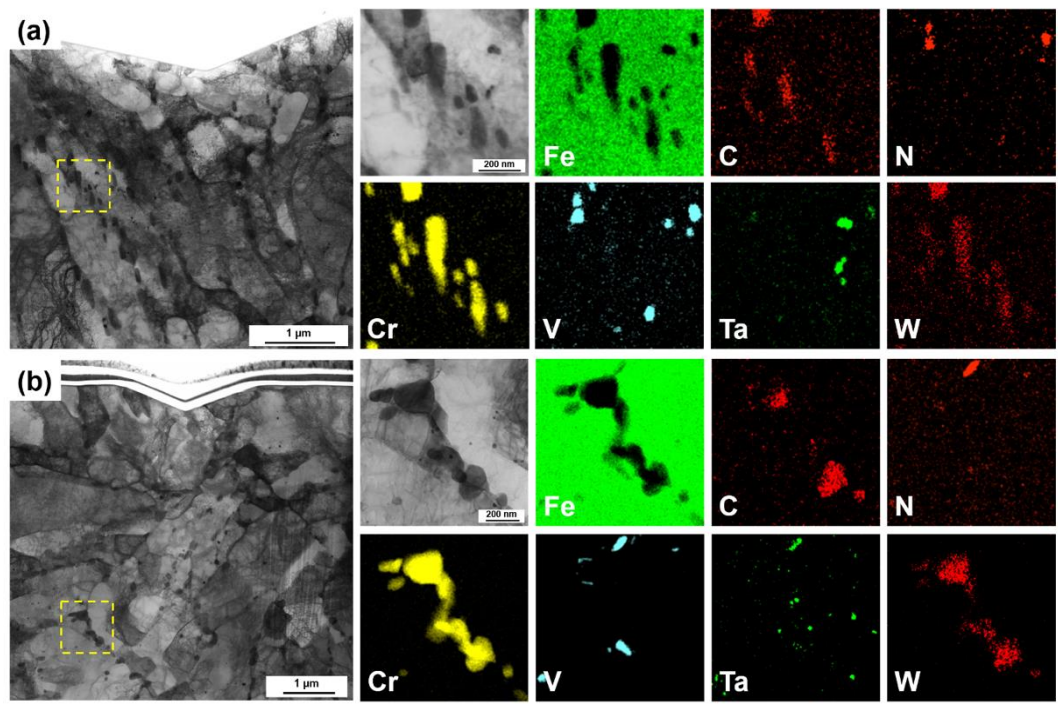

Figure 9. STEM micrographs with the corresponding EDS elemental maps of Eurofer97 after the (a) RT and (b) HT nanoindentation.

Fig. 6 shows areas of the material deformed via indentations suggesting (especially the model alloys) probing material near the grain boundaries. Such action is usually demonstrated on the L-D curve by the presence of the pop-ins, whose magnitude usually increases due to the presence of grain boundaries and may impact recorded mechanical properties. One should remember that the first pop-in can result from plasticity initiation caused by homogeneous dislocation nucleation [56]. Only the following event suggests the presence of an obstacle, precipitate, or grain boundary, which impacts recorded mechanical properties. Our previous studies demonstrated that grain boundaries have higher hardness than the bulk of the material [57]. However, these tests were done with 10x lower loads, thus improving measurement resolution (even then, the recorded hardness increase was about 5% [58]). In addition, Pohl et al. [59] proved that pop-in phenomena appear spontaneously and usually disappear after one, two, or three such events (which is not the case for our studies). Moreover, these events have been recorded at load ranges an order of magnitude smaller than the events recorded in our study, proving that in our study, we deal with the described DSA effect, which is material (chemical composition) dependent. Finally, the shape of the L-D curve when probing material near the grain boundary is different. In this case, dislocation slip and accumulation at a grain boundary can lead to dislocation transfer into or dislocation nucleation within a neighboring grain with a subsequent displacement burst (therefore, probing grain boundary is demonstrated by two interconnected displacement bursts [59, 60]). We never observed such features at our L-D curves, which again points toward DSA. The RT indentation resulted in the formation of dislocation cell structures and forest networks below the indent for all the samples as shown in Fig. 6 and 8, which is commonly observed among similar BCC metals and alloys [61, 62]. The dislocation density of Eurofer97 does not seem to be related to the proximity of the indented surface (see Fig. 6) unlike other materials. Eurofer97 specimen possessed higher dislocation density (see Fig. 6-8 and Table 3) pre-indentation and elongated grains due to the prior hot-rolling during the material manufacture. At HT indentation, Fe, Fe-9Cr, and Fe-9Cr-NiSiP specimens undergo dynamic recovery [62] and thus, result in dislocation substructure polygonization. Increased dislocation mobility at HT contributes to such substructure formation. Polygonization is a well-known recovery mechanism during deformation under high temperatures, where dislocations rearrange to reduce the strain energy [63, 64]. However, Eurofer97 undergoes dynamic recrystallization during HT indentation, resulting in equiaxed grain formation from an initial state of elongated grains. Hence, the stored energy associated with the dislocation density of pre-indentation state acts as the driving force for dynamic recrystallization. Previously, Oliveira et al. [35] reported that the recrystallization of Eurofer97 may occur at 650 °C. At the same time, Stornelli et al. [36] proved that recrystallization can also be observed at lower temperatures, e.g., 600 °C. Our results support that Eurofer97 can recrystallize at 600 °C even after a few hours at this temperature. It has been reported that recrystallization eliminates residual stresses as a result of the nucleation and growth of new defect-free grains within the microstructure of the deformed material [65]. However, the specimen was not fully recrystallized as we still see dislocations within some of the grains, as indicated in Fig. 6g, 7, and 8g. Also, because the quantity of dislocation is greatly reduced, the recrystallization material should have lower strength but higher ductility. Our studies proved this, as we observed lower GND density (see Fig. 8), which confirms this phenomenon and suggests lower material strength, as confirmed by significantly lower hardness and young modulus in all tested specimens, see Fig. 4. This could also be a reason for a lack of DSA effect for these measurement conditions.

Fig. 10 shows the misorientation angle distributions for Fe-9Cr-NiSiP and Eurofer 97 for HT indentation determined using the TKD results described in Fig. 7 via MTEX MATLAB software [66]. The Fe-9Cr-NiSiP specimen shows the higher fraction of low misorientation angle grain boundaries (< 15 degrees) confirming dynamic recovery via polygonization of dislocation substructures. However, the Eurofer97 specimen illustrates a higher fraction of high misorientation angle grain boundaries (> 45 degrees) indicating dynamic recrystallization leading to the nucleation of the defect-free grains.

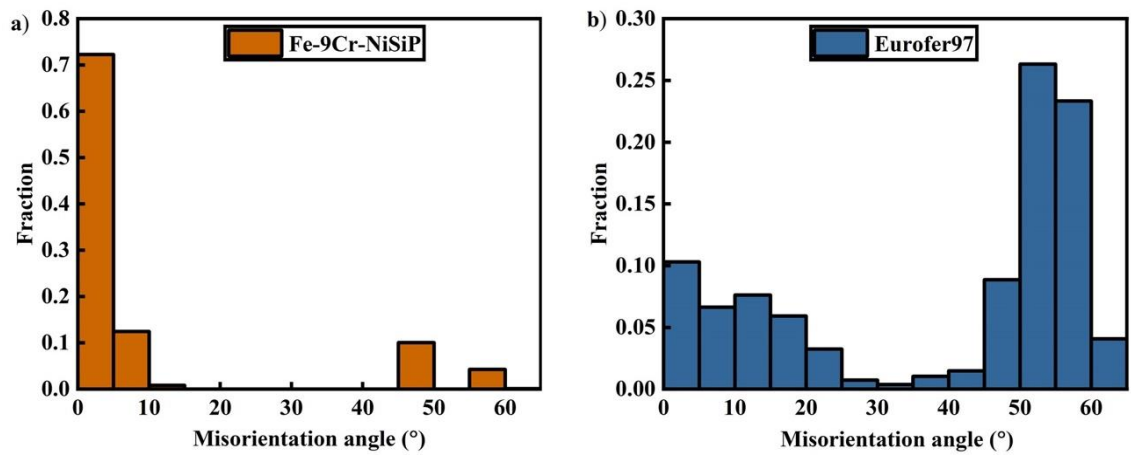

Figure 10. The misorientation angle distribution for Fe-9Cr-NiSiP and Eurofer 97 for HT indentation determined using the TKD as presented in Fig. 7.

Lessons learned and recommendations for future studies:

Despite the constant purging with pure argon, oxidation could be considered as a possible (additional) cause for the bursts. However, the SEM and TEM observations show that after spending ~72h at 600 °C, the oxide layers possess thicknesses of Eurofer97 ~ 100 nm, Fe-9Cr-NiSiP ~ 60 nm, and Fe-9Cr ~ 80 nm (due to sample roughness, the measurement error was about 20%). Furthermore, one would expect to see a confirmation of the presence of an oxide layer at the beginning of the load-displacement curves, which was not the case (development of large pop-in due to oxide cracking). It should also be noted that indentations were made up to 1.5 μm depth. This means the plastic deformation developed under the indenter tip reaches more than 10 μm. This suggests that 60-100 nm thick oxide (recorded after the longest exposure to the highest temperature) plays a minor role when probing such volumes of the material. Please see supplementary material Fig. E and F to evaluate the sample surface after the HT indentation test.

Finally, one should remember that the differences in dislocation observations in TEM between samples (Fig. 6) may be due to material-related differences and where the lamellae were extracted from. Removing a lamella from the region beneath the tip of the indent is difficult, and the lamellae observed could originate from the region between the tip and an edge of the indent (hence the differences in apparent indent depths in the TEM results). This analysis requires further investigation to properly describe the microstructural features developed under the indenter tip.

From lessons gained from these experiments and recent literature [67-71], the following recommendations are provided for future tests. Next experiments are performed at temperatures around the DSA temperature: 150 / 250 °C for pure Fe, and 350 / 450 °C for Fe–9Cr and Fe–9Cr–NiSiP. One sample for each temperature should be tested to avoid annealing of the strain introduced by indentation while performing tests at higher temperatures and avoiding additional oxidation growth on the indents. Modifying experimental conditions to test their influence, such as loading/unloading rates, maximum load (to assess the impact of the probed volume), and holding time (it was found that if the sample was not left to creep enough during the holding time, the modulus results could be strongly impacted [32]). The sample was indented at RT post-HT indentation to measure the impact of any oxide layer present on the surface on the hardness. This would hopefully allow the separation of thermal effects from oxidation layer effects.

## Conclusion

In this work, the evolution of nanoindentation responses at room temperature, 200 °C, 400 °C, and 600 °C, was studied as a function of microstructural complexity increase of different low-activation model materials, i.e., Fe9Cr–1WVTa (Eurofer97), Fe–9Cr–NiSiP, Fe–9Cr alloys, and pure iron. The results are summarized in the following:

1. The high-temperature nanoindentation results show that the hardness and young modulus values do not solely decrease with temperature as expected. This effect is related to the presence of Dynamic Strain Aging phenomena and is strongly correlated with the presence of interstitial and substitutional elements in the metal matrix.

2. According to FIB/TEM analysis of indent cross-sections, nanoindentation deformation occurred through increased dislocation concentration and grain polygonization, while high-temperature nanoindentation resulted in the recrystallization of the samples (especially for the Eurofer97 specimen). This has also been recognized by decreased strains and GND density below the indents, further contributing to the reduction of recorded mechanical properties.

3. High-temperature nanoindentation results are affected by many factors, e.g., dynamic strain aging, and recrystallization. During follow-up high-temperature nanoindentation testing, these factors should be fully considered and analyzed by advanced microscopy.

## **Acknowledgments**

This work was supported by the Euratom research and training programme 2014-2018 under grant agreement No. 755039 (M4F project) and has been supported by the EURATOM Direct Actions. This work also contributes to the Joint Programme on

Nuclear Materials (JPNM) of the European Energy Research Alliance (EERA). We acknowledge support from the European Union Horizon 2020 research and innovation program under NOMATEN Teaming grant agreement No. 857470 and from the European Regional Development Fund via the Foundation for Polish Science International Research Agenda Plus Program Grant No. MAB PLUS/2018/8 and the initiative of the Ministry of Science and Higher Education 'Support for the activities of Centers of Excellence established in Poland under the Horizon 2020 program' under agreement No. MEiN/2023/DIR/3795. Also, financial support from the National Centre for Research and Development through a research grant "Studies of the role of interfaces in multi-layered, coated and composite structures" PL-RPA2/01/INLAS/ 2019 is gratefully acknowledged.